\documentclass{article}
\usepackage[ansinew]{inputenc}
\usepackage{amsmath}
\usepackage{amssymb}
\usepackage{amsthm}
\usepackage{algorithm}
\usepackage{algorithmic}
\usepackage{makeidx}
\usepackage{graphicx}
\usepackage{cite}
\newtheorem{probb}{Problem}

\newtheorem{theo}{Theorem}
\newtheorem{corr}{Corollary}
\newtheorem{lemm}{Lemma}
\newtheorem{propp}{Proposition}

\theoremstyle{definition}

\theoremstyle{remark}

\begin{document}
\title{Efficient algorithms for deciding the type of growth of products
of integer matrices}
\author{Rapha\"el Jungers \thanks{R. Jungers and V. Blondel are with the Department of
Mathematical Engineering, Universit\'e catholique de Louvain,
Avenue Georges Lemaitre 4, B-1348 Louvain-la-Neuve, Belgium,
{\tt\small \{jungers, blondel\}@inma.ucl.ac.be}. Their research
was partially supported by Communaut\'e francaise de Belgique -
Actions de Recherche Concert\'ees, by the HYCON Network of
Excellence (contract number FP6-IST-511368) and by the Belgian
Programme on Interuniversity Attraction Poles initiated by the
Belgian Federal Science Policy Office. The scientific
responsibility rests with its authors. Rapha\"el Jungers is a FNRS
fellow (Belgian Fund for Scientific Research).}, Vladimir Protasov\thanks{ V.Protasov is with the department of Mechanics and
Mathematics, Moscow State University, Vorobyovy Gory, Moscow,
119992, Russia, {\tt\small vladimir\_ protassov@yahoo.com}. His
research is supported by the grant RFBR 05-01-00066 and by the
grant 5813.2006.1 for the Leading Scientific Schools. }, Vincent D. Blondel}
%
%

\maketitle
\begin{abstract}
For a given finite set $\Sigma$ of matrices with nonnegative
integer entries we study the growth of $$\, \max_t(\Sigma ) \, =
\, \max\{\|A_{1}\dots  A_{t}\|: A_i  \in \, \Sigma\}.$$ We show
how to determine in polynomial time whether the growth with $t$ is
bounded, polynomial, or exponential, and we characterize precisely
all possible behaviors.

\end{abstract}
\begin{section}{Introduction}

 In the last decade the joint spectral radius of sets of matrices has
been the subject of intense research due to its role for studying
wavelets, switching systems, approximation algorithms, curve
design, {etc.} \cite{blne-appr-jsr,Rio,Ref}. The particular case
of integer (rather than real) matrices is itself interesting due
to the existence of many applications where such matrices arise.
For instance, the rate of growth of the binary partition function
in combinatorial number theory is expressed in terms of the joint
spectral radius of binary matrices, that is, matrices whose
entries are zeros and ones \cite{R,P3}. Moision et al.
\cite{moision-joint-constraints,mos-bounds,moision01codes} have
shown how to compute the capacity of a code under certain
constraints (caused by the noise in a channel) with the joint
spectral radius of binary matrices. Recently the joint spectral
radius of binary matrices has also been used to express
trackability of mobiles in a sensor network \cite{cyb}.

For a given finite set $\Sigma$ of matrices, the \emph{joint
spectral radius} of the set $\Sigma$, denoted $\rho(\Sigma)$, is
defined by the limit $$\rho(\Sigma) = \lim_{t \to
\infty}{\max_t(\Sigma )^{1/t}}= \lim_{t\to \infty}
\max\{\|A_{1}\dots  A_{t}\|^{1/t}: A_i  \in \, \Sigma\}.$$This
limit exists for all finite sets of matrices and does not depend
on the chosen norm. In the sequel we will mostly use the norm
given by the sum of the absolute values of all matrix entries. Of
course, for nonnegative matrices this norm is simply given by the
sum of all entries.

The  problem of computing the joint spectral radius is known to be
algorithmically undecidable in the case of arbitrary matrices.
There are several known approximation algorithms
\cite{blne-appr-jsr, P1,P2,CH}, but all of them have exponential
complexity either in the dimension of the matrices or in the
accuracy of computation. Even in the case of binary matrices,
computing the joint spectral radius is not easy: that problem has
been shown to be NP-hard \cite{blondel98survey}.

In this paper, we focus on the case of \emph{nonnegative integer}
matrices and consider questions related to the growth with $t$ of
$\max_t (\Sigma)$.  When the matrices are nonnegative all the
following cases can possibly occur:

\begin{enumerate}
\item $\rho(\Sigma)=0$. Then $\max_t (\Sigma)$ takes the value 0
for all values of $t$ larger than  some $t_0$ and so all products
of length at least $t_0$ are equal to zero. \item $\rho(\Sigma)=1$
and the products of matrices in $\Sigma$ are bounded, that is,
there is a constant $K$ such that $\|A_{1} \dots {A_{t}}\| <K$ for
all $A_{i} \in \Sigma$. \item $\rho(\Sigma)=1$ and the products of
matrices in $\Sigma$ are unbounded. We will show in this
contribution that in this case the growth of $\max_t (\Sigma)$ is
polynomial. \item $\rho(\Sigma)>1$. In this case the growth of
$\max_t (\Sigma)$ is exponential.
\end{enumerate}

Note that
 the situation $0 < \rho(\Sigma)<1$ is not possible because the norm of
a nonzero integer matrix is always larger than one. The  cases (1)
to (4) already occur when there is only one matrix in the set
$\Sigma$. Particular examples for each of these four cases are
given by the matrices:
\begin{equation*}
\left( \begin{array}{cc} 0 & 1 \\ 0 & 0
\end{array}\right), \; \left( \begin{array}{cc} 1 & 0 \\ 0 & 1
\end{array}\right), \; \left( \begin{array}{cc} 1 & 1 \\ 0 & 1
\end{array}\right), \; \left( \begin{array}{cc} 1 & 1 \\ 1 & 1
\end{array}\right).
\end{equation*}

 The problem of distinguishing
between the different cases has a long history. The
polynomial-time decidability of the equality $\rho(\Sigma)=0$ is
shown in \cite{gu}. As mentioned by Blondel and Canterini
\cite{blon-cant}, the decidability of the boundedness of products
of nonnegative integer matrices follows from results proved in the
$70$s. Indeed, the finiteness of a semigroup generated by a finite
set of  matrices has been proved to be decidable independently by
Jacob \cite{jacob} and by Mandel and Simon \cite{man-sim}. It is
clear that for integer matrices, finiteness of the semigroup is
equivalent to its boundedness, and so boundedness is decidable for
integer matrices. The decision algorithms proposed in \cite{jacob}
and \cite{man-sim} are based on the fact that if the semigroup is
finite, then every matrix in the semigroup can be expressed as a
product of length at most $B$ of the generators, and the bound $B$
only depends on the dimension of the matrices $n$ and on the
number of generators. The proposed algorithms consist of
generating all products of length less than $B$; and checking
whether new products are obtained by considering products of
length $B+1$. The high value of the bound $B$ does however lead to
highly non polynomial algorithms and is therefore not practical. A
sufficient condition for the unboundedness of $\max_t (\Sigma)$
was also derived  recently for the case of binary matrices by
Crespi et al. \cite{cyb}. We show here that the condition given
there is also necessary. Moreover, we provide a polynomial
algorithm that checks this condition, and thus we prove
 that boundedness of semigroups of integer matrices is  decidable in
polynomial time.
  Crespi et al. \cite{cyb} also provide a criterion to
verify the inequality $\rho (\Sigma ) > 1$ for binary matrices and
an algorithm based on that criterion. However, their algorithm is
not polynomial\footnote{The comments made here on the results
presented in the Technical Report \cite{cyb} refer to the version
of August 11, 2005 of that report. In a later version,  and after
a scientific exchange between RJ and VB with two of the authors of
the report, the authors of \cite{cyb} have improved some of their
results and have incorporated some of the suggestions made by RJ
and VB, as acknowledged in the updated version of the report dated
 December 19, 2005.}. In this paper, we present a polynomial algorithm
for checking $\rho (\Sigma ) > 1$ for sets of nonnegative integer
matrices. Let us note that the same problem for other joint
spectral characteristics (such as the lower spectral radius or the
Lyapunov exponent) is proved to be NP-hard even for binary
matrices \cite{tsitsiklis97lyapunov}. Therefore, the polynomial
solvability of this question for the joint spectral radius is
somewhat surprising.

Our results have direct implications for all the problems that can
be formulated in terms of a joint spectral radius of nonnegative
integer matrices. In particular, it follows from our results that
the trackability problem for sensor networks as formulated in
\cite{cyb} can be decided in polynomial time. The trackability
problem is as follows: we are given a directed graph with labelled
nodes. Nodes may have identical labels and we consider successions
of labels produced by directed paths in the graph. The function
$N(t)$ gives the largest number of paths that are compatible with
some label sequence of length $t$. When the growth of $N(t)$ is
bounded or grows polynomially, the graph is said to be trackable.
It has been shown in \cite{cyb} that trackability can be decided
by verifying that the joint spectral radius of two binary matrices
constructed from the graph is less or equal to one. In this paper,
this last property is shown to admit a polynomial time decision
algorithm. Moreover, we provide an algorithm for computing the
degree of the polynomial growth for trackable graphs.

Our main results can be summarized as follows. For any finite set
of nonnegative integer $n\times n$ matrices $\Sigma $ there is a
polynomial algorithm that decides between the four cases $\rho=0$,
$\rho=1$ and bounded growth, $\rho=1$ and polynomial growth, $\rho
>1$ (see Theorem 1 and Theorem 2). Moreover, if $\rho(\Sigma)=1$,
then there exist constants $C_1,C_2,k$, such that
 $\, C_1t^k\, \le \, \max_t (\Sigma) \, \le \, C_2 t^k$ for all $t$;
the rate of growth $k$ is
 an integer such that $0 \leq k \leq n-1$, and there is
  a polynomial time algorithm for computing $k$ (see Theorem 3).
This sharpens previously known results on the asymptotic of the
value $\max_t (\Sigma)$ for nonnegative integer matrices. We
discuss this aspect in Section 6.
  Thus,
for nonnegative integer matrices, the only case for which we
cannot decide the exact value of the joint spectral radius is
$\rho > 1$. However, it is most likely that the joint spectral
radius cannot be polynomially approximated in this case since it
was proved that its computation is NP-hard, even for binary
matrices \cite{blondel98survey,tsitsiklis97lyapunov}.

 The paper is
organized as follows. Section \ref{tools} contains some notation
and auxiliary facts from graph theory. In Section
\ref{section-behaviours} we establish a criterion for separating
the three main cases  $\rho (\Sigma ) < 1, \rho (\Sigma ) = 1$ and
$\rho (\Sigma ) > 1$. Applying this criterion we derive a
polynomial algorithm that decides each of these cases. In Section
\ref{bound} we present a criterion for deciding product
boundedness  and provide a polynomial time implementation of this
criterion. In Section \ref{poly} we find the asymptotic behavior
of the value $\max_t (\Sigma )$ as $t \to \infty$ for the case
$\rho = 1$. We prove that this value is asymptotically equivalent
to $t^k$ for a certain integer $k$ with $0 \leq k \leq n-1$ and
show how to find the rate of growth $k$ in polynomial time.
Finally, in Section \ref{problems} we formulate several open
problems on possible generalizations of those results to arbitrary
matrices.\end{section}


\begin{section}{Auxiliary facts and notation} \label{tools}

For a given finite set of matrices $\Sigma$ we denote by
$\Sigma^t$ the set of all products of length $t$ of matrices from
$\Sigma$. By $\Sigma^{\star}$ we denote the union of all
$\Sigma^t$ over all $t \geq 0$. For two nonnegative functions
$f(t), g(t)$ we use the standard notation $f(t ) = O(g(t))$, which
means that there is a positive constant $C$ such that $f(t) \le
Cg(t)$ for all $t$. The functions $f$ and $g$ are said to be
asymptotically equivalent, which we denote $f(t) \asymp g(t)$ if
$f(t ) = O(g(t))$ and $g(t ) = O(f(t))$.

We shall consider each nonnegative $n\times n$ matrix as the
adjacency matrix of a directed weighted graph $G$. This graph has
$n$ nodes enumerated from $1$ to $n$. There is an edge from node
$i$ to node $j$ if the $(i,j)$ entry of the  matrix is positive
and the weight of this edge is then equal to the corresponding
entry. This graph may have loops, {i.e.}, edges from a node to
itself, which correspond to diagonal entries. If we are given a
family $\Sigma$ of nonnegative integer matrices, then we have
several weighted graphs on the same set of nodes $\{1, \ldots ,
n\}$. In addition we define the graph $G(\Sigma)$ associated to
our family $\Sigma$ as follows: There exists an edge in
$G(\Sigma)$ from node $i$ to node $j$ if and only if there is a
matrix $A \in \Sigma$ such that $A_{i, j} > 0$. The weight of this
edge is equal to $\max\limits_{A \in \Sigma} A_{i, j}$. We shall
also use the graph $G^2$, whose $n^2$ nodes represent the ordered
pairs of our initial $n$ nodes, and whose edges are defined as
follows: there is an edge from a node $(i,i')$ to $(j, j')$ if and
only if there is a matrix $A \in \Sigma$ such that both $A_{i, j}$
and $ A_{i', j'}$ are positive \emph{for the same matrix}. The
edges of $G^2$ are not weighted.

 Products of matrices from
$\Sigma$ can be represented by \emph{cascade graphs}. In a cascade
graph, a matrix $A \in \Sigma$ is represented by a bipartite graph
with a left and a right set of nodes. The sets have identical size
and there
 is an edge between the $i$th left node
and the $j$th right node if $A_{i,j}>0$. The weight of this edge
is equal to the entry $A_{i,j}$. For instance, the non-weighted
bipartite graph on Figure \ref{fig-bip} represents the matrix
$$\begin{pmatrix} 1&1&0\\ 0&0&1\\0&0&1 \end{pmatrix}.$$
\begin{figure}
 \centering
\includegraphics[width=0.3\textwidth ]{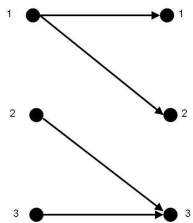}

 \caption{A bipartite graph representing a binary matrix.}
\label{fig-bip}
 \end{figure}

\begin{figure}
 \centering
\includegraphics[width=0.3\textwidth ]{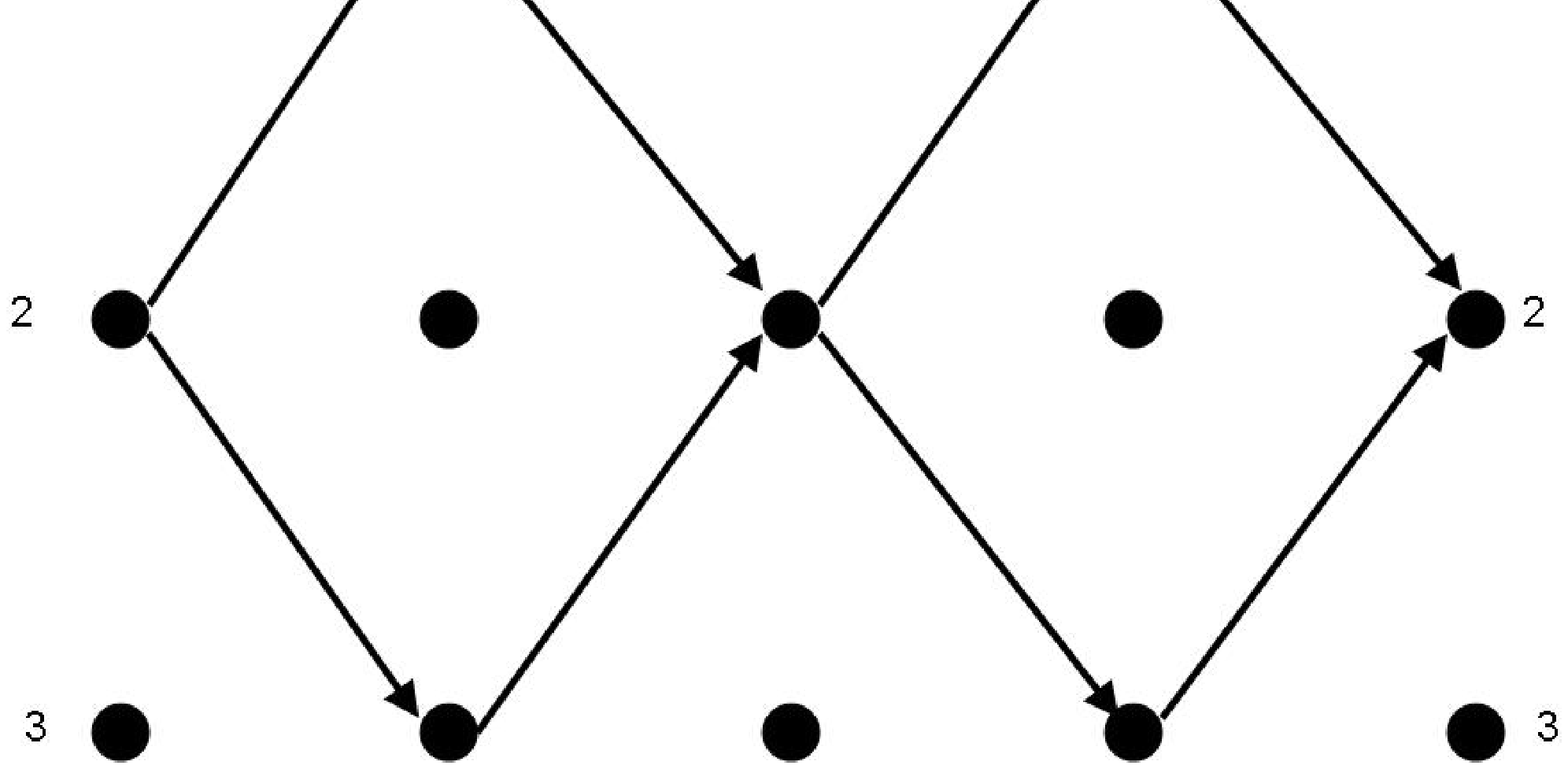}
 \caption{A typical cascade graph.}
 \label{fig-expo} \end{figure}

Now, for a given product of matrices $A_{d_{1}}\dots A_{d_{t}}$ we
construct a cascade graph as follows: we concatenate the
corresponding bipartite graphs in the order in which they appear
in the product, with the right side of each bipartite graph
directly connected to the left side of the following graph. For
example, Figure \ref{fig-expo}  shows a cascade graph representing
 the product $A_0A_1A_0A_1$ of length four, with
$$A_0=\begin{pmatrix} 0&0&0\\ 1&0&1\\0&0&0
\end{pmatrix}, A_1=\begin{pmatrix} 0&1&0\\ 0&0&0\\0&1&0
\end{pmatrix}.$$

We say that the bipartite graph at the extreme left side begins at
level $t=0$ and the one at the extreme right side ends at the last
level.  We note $(i,t)$ to refer to the node $i$ at level $t$.  We
say that there is a path from node $i$ to node $j$ if one is able
to construct a cascade graph with a path from some node $(i,t)$ to
some node $(j,t')$ for some $t<t'.$  A path is to be understood as
a succession of edges from a level to the next level, {i.e.}
always from left to right. One can check that the $(i,j)$ entry of
a matrix product of length $t$ is equal to the number of directed
paths from the node $(i,0)$ to the node $(j,t)$ in the
corresponding cascade graph.  We thus have a way of representing
$\max_t (\Sigma)$ as the maximal total number of paths from
extreme left nodes to extreme right
nodes in cascade graphs of length $t$.  

Two nodes of a graph are called connected if they are connected by
a path (not necessarily by an edge). A directed graph is strongly
connected if for any pair of nodes $(i,j)$, $i$ is connected to
$j$. The following well known result states that we can partition
the set of nodes of a directed graph in a unique way in strongly
connected components, and that the links between those components
form a tree \cite{tarjan}. \\

\begin{lemm}\label{l0}
For any directed graph $G$ there is a partition of its nodes in
nonempty  disjoint sets $V_1, \ldots , V_I$ that are strongly
connected and such that no two nodes belonging to different
partitions are connected by directed paths in both directions.
Such a maximal decomposition is unique up to renumbering. Moreover
there exists a (non necessarily unique) ordering of the subsets
$V_s$ such that any node $i \in V_k$ cannot be connected to any
node $j \in V_l$, whenever $k
> l$. There is an algorithm to obtain this partition in $O(n)$
operations (with $n$ the number of nodes).
\end{lemm}
%
In this lemma, we suppose by convention that a node that is not
strongly connected to any other node is itself a strongly
connected subset, even if it does not have a self-loop. In such a
case we will say that the corresponding set is a \emph{trivial
strongly connected subset}.
 Consider the graph
$G(\Sigma)$ corresponding to a family of matrices $\Sigma$, as
defined above. After possible renumbering, it can be
assumed that the set of nodes is ordered, that is, for all nodes
$i \in V_k$ and $j \in V_l$, if $k
>l$ then $i>j$. In that case all the matrices of $\Sigma$ have block
upper-triangular form with $I$ blocks corresponding to the sets
$V_1, \ldots ,
V_I$ ($I$ can be equal to one). 
\end{section}


\begin{section}{Deciding
$\, {\mathbf \rho <1}\, $, $\, {\mathbf \rho =1},\, $ and $\,
{\mathbf \rho >1}$.}\label{section-behaviours}

Let $\Sigma$ be a finite set of nonnegative integer matrices and
let $\rho = \rho (\Sigma )$ be their joint spectral radius. The
goal of this section is to prove the following result.\\

\begin{theo}\label{th1}
For matrices with nonnegative integer entries there is a
polynomial algorithm that decides the cases $\rho < 1, \rho = 1$
and $\rho
>1$.\end{theo}
\begin{proof}
The proof will be split into several lemmas. The inequality $\rho
< 1$ means that the maximum number of paths in a cascade graph of
length $t$ tends to zero as $t \to \infty$.  Hence for
sufficiently large $t$ there are no paths of this length in the
graph $G(\Sigma)$ corresponding to the whole family $\Sigma$,
since this graph represents the set of all possible edges. This
means that $G(\Sigma)$ has no cycles. So we get our first lemma :

\begin{lemm}\label{cond-zero}
For a finite set of nonnegative integer matrices $\Sigma$, we have
$\rho (\Sigma)>0$ if and only if the graph~$\, G(\Sigma)$ has a
cycle. In this case $\rho \ge 1$.
\end{lemm}
This condition can be checked in $O(n)$ operations : one just has
to find the strongly connected components of the graph $G(\Sigma)$
(a task that can be performed in $O(n)$ operations \cite{tarjan});
a cycle will be possible iff one of the subsets is non trivial.
The problem of deciding between  $\rho=1$ and  $\rho >1$ is more
difficult. Let us start with the following lemma.\\

\begin{lemm}\label{lemdiag}
Let $\Sigma$ be an arbitrary finite set of real matrices. If $\rho
(\Sigma) > 1$, then there is a product  $A \in \Sigma^{\star}$,
for which $A_{i, i }>1$ for some~$i$. If the matrices are
nonnegative, then the converse is also true.
\end{lemm}
\begin{proof}
\textbf{Sufficiency}. Since $A \in \Sigma^t$ has nonnegative
elements, it follows that $\|A^k\| \ge A_{i, i}^k$, hence $\rho
(A) \ge A_{i, i} > 1$. It is well-known that for all $t$, and for
all $A\in \Sigma^t$, $\rho(\Sigma)\geq \rho(A)^{1/t}$; therefore
$\rho (\Sigma )\ge [\rho (A)]^{1/t} >1$.\\ \textbf{Necessity}.
Since $\rho (\Sigma ) >1$ it follows that there is a product $B
\in \Sigma^{\star}$ such that $\rho (B) > 1$ \cite{BW}. Let
$\lambda_1$ be one eigenvalue of $B$ of largest magnitude, so
$|\lambda_1| = \rho (B) > 1$ and let $\lambda_2, \ldots ,
\lambda_n$ be the other eigenvalues. There exists a $t$
sufficiently large such that $|\lambda_1|^t > 2n$ and $\arg (
\lambda_k^t) \in \bigl(-\frac{\pi}{3}, \frac{\pi}{3} \bigr)$
 for all $k = 1, \ldots , n$, where $\arg ( z)$ is the argument of the
complex number $z$
  \cite{vinogradov}. Therefore ${\rm Re}( \lambda_k^t ) \ge \, \frac12
|\lambda_k^t|$ for all $k$. We have  $ \sum\limits_{k=1}^n
(B^t)_{k,k}  =  {\rm tr}\, B^t  =
 \sum\limits_{k=1}^n \lambda_k^t  =  \sum\limits_{k=1}^n {\rm Re}\,
\lambda_k^t \ge \frac{1}{2}|\lambda_1^t| > n$. Since the sum of
the $n$ numbers $(B^t)_{k,k}$ exceeds $n$, hence one of them must
exceed $1$.
\end{proof}

\begin{corr}\label{cordiag}
For any finite set  of nonnegative integer matrices $\Sigma$, we
have
 $\rho (\Sigma) > 1$ if and only if
there is a product $A \in \Sigma^{\star}$ such that $A_{i, i } \ge
2$ for some~$i$. \end{corr} A different proof of this corollary
can be found in Crespi et al. \cite{cyb}. Thus, the problem is
reduced to testing if there is a product $A \in \Sigma^{\star}$
that has a diagonal element larger or equal to $2$. This is
equivalent to the requirement that at least one of the following
conditions is satisfied:
\begin{enumerate}
\item There is a cycle in the graph $G(\Sigma)$ containing at
least one edge of weight greater than $2$. \item There is a cycle
in the graph $G^2$ containing at least one node $(i,i)$ (with
equal entries) and at least one node $(p,q)$ with $p \ne q$.
\end{enumerate}
Indeed, if $A_{i,i} \ge 2$ for some $A\in \Sigma^{\star}$, then
either there is a path on the graph $G(\Sigma)$ from $i$ to $i$
that goes through an edge of weight $\ge 2$ (first condition), or
there are two different paths from $i$ to $i$ in the cascade graph
corresponding to the product $A$, this is equivalent to the second
condition. The converse is obvious. To verify Condition 1 one
needs to look over all edges of $G(\Sigma)$ of weight $\ge 2$ and
to check the existence of a cycle containing this edge. This
requires at most $n^3$ operations. To verify Condition 2 one needs
to look over all $\frac{1}{2}n^2(n-1)$ triples $(i, p, q)$ with
$p>q$ and for each of them check the existence in the graph $G^2$
of paths from $(i, i)$ to $(p,q)$ and from  $(p,q)$ to $(i, i)$,
which requires at most $n^2$ operations. Thus, to test Condition 2
one needs to perform at most $n^5$ operations. This completes the
proof of Theorem \ref{th1}.\end{proof}

Figure \ref{fig-expo} shows a cascade graph with the condition 2
of Corollary 1 satisfied: there are two paths from node $2$ to
node $2$, and for every even $t$, the number of paths is
multiplied by two.

The shortest cycle in the graph $G^2$ with the required properties
has at most $n^2$ edges. It therefore follows that whenever $\rho
(\Sigma)
> 1$, there is a product $A$ of length less than $ n^2$ such that
$A_{i, i } \ge 2$ for some $i$. From this we deduce the following
corollary.\\

\begin{corr}\label{corjsr}
Let $\Sigma$ be a finite set of nonnegative integer matrices of
dimension $n$. If $\rho (\Sigma)
> 1$, then $\rho (\Sigma) \ge 2^{1/n^2}$.
\end{corr}

\end{section}


\begin{section}{Deciding product boundedness}\label{bound}
If $\rho =1$, two different cases are possible: either the maximum
norm of products of length $t$ is bounded by a constant, or it
grows with $t$. Deciding between these two cases is not trivial.
In this section we present a simple criterion that allows us to
decide whether the products are bounded. Our reasoning will be
split into several lemmas. We begin with a simple but crucial
observation.\\

\begin{lemm}\label{lemreduc}
Let $\Sigma$ be a finite set of nonnegative integer  matrices with
${\rho(\Sigma ) = 1}$. If there is a product $A \in
\Sigma^{\star}$ that has an entry larger than $1$, then the graph
$G(\Sigma)$ is not strongly connected.\end{lemm}
\begin{proof}
Let  $A_{i, j} \ge 2$, that is, counting with weights, there are
two paths from $i$ to $j$ in the same cascade graph.  If there is
another cascade graph with a path from $j$ to $i$, then,
concatenating the two cascade graphs, we can find two different
paths from $i$ to itself, and by corollary \ref{cordiag} $\rho
(\Sigma ) >1$, which is a contradiction.
Hence $G(\Sigma)$ is not strongly connected.\end{proof}
Consider the partition of the nodes of $G(\Sigma)$ into
strongly connected sets $V_1, \ldots , V_I$ (\emph{cfr.} Lemma
\ref{l0}). Applying Lemma \ref{lemreduc} we get the following
corollaries.\\

\begin{corr}\label{correduc}
Let $\Sigma$ be a finite set of nonnegative integer  matrices. If
$\rho (\Sigma)=1$, but the products of these matrices are not
uniformly bounded, then  there exists a permutation matrix $P$
such that for all matrix $A$ in $\Sigma$, $P^TAP$ is block upper
triangular with at least two blocks.\end{corr}

\begin{corr}\label{corones}
Let $\Sigma$ be a finite set of nonnegative integer  matrices with
joint spectral radius one. Then all products of those matrices
restricted to any strongly connected set $V_k$ are binary
matrices.
\end{corr}

We are now able to prove the main result of this section. We first
provide a result for the case of one matrix and then consider the
case of several matrices.\\

\begin{propp}\label{lem-unbounded}
Let $A$ be a nonnegative integer matrix with $\rho(A)=1$.  The set
$\{\|A^t\|: t\geq 1\}$ is unbounded if and only if there exists
some $k\geq 1$, and a pair of indices $(i,j)$ such that
\begin{equation}
A^k_{i,i},
 A^k_{i,j}, A^k_{j,j} \geq 1.
\end{equation}
\end{propp}
\begin{proof}
Sufficiency is easy: One can check that  $(A^{kt})_{i,j} \ge t$
for any $t$, and hence $\max_t (\Sigma)$ is unbounded.  Let us
prove the necessity : Consider the partition in strongly connected
subsets $V_1, \ldots , V_I$. By Corollary \ref{correduc} we have
$I \ge 2$.\\
 We claim that there are two
nontrivial sets $V_a$ and $V_b , \, a < b$ that are connected by a
path (there is a path from an element of $V_a$ to an element of
$V_b$). Otherwise any path in $G(\Sigma)$ intersects at most one
nontrivial set, and we prove that their number must then be
bounded : Let a path start from a set $V_{a_1}$, then go to
$V_{a_2}$ etc., until it terminates on $V_{a_l}$. We associate the
sequence $a_1 < \cdots < a_l,\ l \le I$  to this path.  As
supposed, this sequence contains at most one nontrivial set, say
$V_{a_s}$. There are at most $K^{l}$ paths, counting with weights,
corresponding to this sequence, where $K$ is the largest number of
edges between two given sets (still counting with weights).
Indeed, each path of length $t > l$ begins with the only edge
connecting $V_{a_1}$ to $V_{a_2}$ (since $V_{a_1}$ is trivial),
\emph{etc.} until it arrives in $V_{a_s}$ after $s-1$ steps (for
each of the previous steps we had at most $K$ variants), and the
reasoning is the same if one begins by the end of the path, while,
given a starting node in $V_{a_s}$, and a last node in the same
set, there is at most one path between these two nodes, by
corollary \ref{corones}.
Since
there are finitely many  sequences $\{a_j\}_{j=1}^l, \ l \le I$,
we see that the total number of paths of length $t$ is bounded by
a constant independent of $t$, which contradicts the assumption.\\
Hence there are two nontrivial sets $V_a$ and $V_b$, $a < b$
connected by a path. Let this path go from a node $i_1 \in V_a$ to
$j_1 \in V_b$ and have length $l$. Since both graphs $V_a$ and
$V_b$ are strongly connected, it follows that there is a cycle
$i_1 \to \ldots \to i_p \to i_1$ in $V_a$ and a path $j_1 \to
\ldots \to j_q \to j_1$ in $V_b,\ p,q \ge 1$.    Take now a number
$s \in \{1, \ldots , p\}$ such that $l+s$ is divisible by $p: \
l+s = vp,\ v \in \mathbb N$. Take a nonnegative integer  $x$ such
that $v + x$ is divisible by $q: \ v + x = uq,\ u \in \mathbb N$.
Let us show that the matrix $A^{upq}$ and the indices $i =
i_{p-s+1}, j = j_1$ possess property \ref{unbound}. Indeed, a path
of length $upq$ along the first cycle, beginning at node
$i_{p-s+1}$ terminates to the same node, hence $A^{upq}_{i_{p-s+1}
, i_{p-s+1}} \ge 1$. Similarly $(A^{upq})_{j_1 , j_1} \ge 1$. On
the other hand, the path going from $i_{p-s+1} \to \ldots \to
i_1$, then going x times around the first cycle from $i_1$ to
itself, and then going from $i_1$ to $j_1$, has a total length $s
+ xp + l = vp+xp = upq$, therefore $A^{upq}_{i_{p-s+1} , j_1} \ge
1$.\end{proof}

The fact that there must be two nontrivial sets connected by a
path had already been proved by Mandel and Simon \cite[Lemma
2.6]{man-sim}. We now provide a generalization of this result to
the case of several matrices.\\

\begin{propp}\label{prop-unbounded}
Let $\Sigma$ be a finite set of integer nonnegative matrices with
$\rho(\Sigma)=1$.  The set of products norms $\{ \|A\|: A \in
\Sigma^{\star}\}$ is unbounded if and only if there exists a
product $A \in \Sigma^{\star}$, and indices $i$ and $j$ ($i \neq
j$) such that
\begin{equation}\label{unbound}
A_{i,i}, A_{i,j}, A_{j,j} \geq 1.
\end{equation} \end{propp}
\begin{proof}
The sufficiency is obvious by the previous lemma. Let us prove the
necessity.  We have a set $\Sigma$ of nonnegative integer
matrices, and their products in $\Sigma^{\star}$ are unbounded.
Consider again the partition of the nodes in strongly connected
sets $V_1, \ldots , V_I$ for $\Sigma$. Our proof proceeds by induction on $I$. For $I = 1$ the products are
bounded by corollary \ref{corones}, and there is nothing to prove.
Let $I \ge 2$ and the theorem holds for any smaller number of sets
in the partition. If on the set $U = \cup_{s=2}^I V_s$ the value
$\max_t (\Sigma , U)$ is unbounded, then the theorem follows by
induction. Suppose then that the products are bounded on this
subset of nodes, by some constant $M$.  Let us consider a product
of $t$ matrices, and count the paths from any leftmost node to any
rightmost node.  First, there are less than $n^2$ paths beginning
in $V_1$ and ending in $V_1$, since the corresponding adjacency
matrix must have $\{0,1\}$ entries (recall that $n$ is the total
number of nodes). Second, there are at most $Mn^2$ paths beginning
and ending in $U$, since each entry is bounded by $M$.  Let us
count the paths beginning in $V_1$ and ending in $U$ : Let $i_0
\to \cdots \to i_{t}$ be one of these paths. The nodes $i_0,
\ldots i_{r-1}$ are in $V_1$, the nodes $i_{r}, \ldots , i_t$ are
in $U$ and $i_{r-1}i_r$ is an edge connecting $V_1$ and $U$.  The
number $r$ will be called a switching level. For any switching
level there are at most $KMn^2$ different paths connecting $V_1$
with $U$, where $K$ is the maximum number of edges jumping from
$V_1$ to $U$ at the same level, counting with weights.  Indeed for
one switching edge $i_{r-1}i_r$, the total number of paths from
$i_r$ to any node at the last level is bounded by $M$, and there
are less than $n$ nodes in $U$. By the same way of thinking, there
is maximum one path from each node in $V_1$ to $i_{r-1}$, and
there are less than $n$ nodes in $V_1$.  The number of switching
levels is thus not bounded, because so would be the number of
paths. To a given switching level $r$ we associate a triple $(A',
A'', d)$, where $A' = A_{d_1}\dots {A_{d_{r-1}}}_{\bigl| V_1}$ and
$A'' = A_{d_{r+1}}\dots {A_{d_{t}}}_{\bigl| U}$ are matrices and
$d = d_r$ is the index of the $r$th matrix.  The notation
$A_{\bigl| V_1}$ means the square submatrix of $A$ corresponding
to the nodes in $V_1$. Since $A'$ is a binary matrix (Corollary
\ref{corones}), $A''$ is an integer matrix with entries less than
$M$, and $d$ can take finitely many values, it follows that there
exist finitely many, say $N$, different triples $(A', A'', d)$.
Taking $t$ large enough, it can be assumed that the number of
switching levels $r \in \{2, \ldots , t-1\}$ exceeds $N$, since
for any switching level there are at most $KMn^2$ different paths.
Thus, there are two switching levels $r$ and $r+s, \ s \ge 1$ with
the same triple. Define $d = d_{r} = d_{r+s}$ and
\begin{equation}\label{decompose}
B \ = \ A_1\dots A_{d_{r-1}}\, , \quad D = A_{d_{r+1}} \dots
A_{d_{r+s-1}}\, , \quad E = A_{d_{r+s+1}} \dots A_{d_t}
\end{equation}
(if $s=1$, then $D$ is the identity matrix). Thus, $A_{d_1}\dots
A_{d_t} = BA_dDA_dE$.  Since $A' = B_{\bigl| V_1}\, = \,
BA_dD_{\bigl| V_1}$ it follows that $B_{\bigl| V_1}=
B(A_dD)^k_{\bigl| V_1}$ for any $k$. Similarly $A'' = E_{\bigl| U}
= DA_dE_{\bigl| U}$ implies that  $E_{\bigl| U} =
(DA_d)^kE_{\bigl| U}$. Therefore for any $k$ the cascade graph
corresponding to the product $B(A_dD)^kA_dE$ has at least $k+1$
paths of length $t + (k-1)s$ starting at $i_0$. Those paths have
switching levels $r, r+s, \ldots , r+ (k-1)s$ respectively.
Indeed, for any $l \in \{0, \ldots , k\}$ there is a path from
$i_0$ to ${i_{r-1+ls}} = i_{r-1}$, because $ B(A_dD)^l_{\bigl|
V_1}=B_{\bigl| V_1}$;  there is an edge from ${i_{r-1+ls}}$ to
${i_{r+ls}}=i_r$, because $A_{d_{r+ls}} = A_{d_r} = A_d$; finally
there is a path from ${i_{r+ls}} = i_r$ to $i_{t + (k-1)s} = i_t$,
because $ (DA_d)^{k-l}E_{\bigl| U}=E_{\bigl| U} $. Therefore,
$\|B(A_dD)^kA_dE\| \ge k+1$ for any $k$, hence $\|B(A_dD)^kA_dE\|
\to \infty $ as $k \to \infty$, and so $\|(A_dD)^k\| \to \infty$.
Now we apply the first part of the proof for the matrix $A_dD$; since the powers of this matrix are unbounded it follows
that some power $A = (A_dD)^k$, which is $(A_{d_r} \dots
A_{d_{r+s-1}})^k$ possesses the property $A_{i,i}, A_{j,j},
A_{i,j}\geq 1$ for suitable $i$ and $j$.\end{proof}

There is also a different way to derive Proposition
\ref{prop-unbounded}. Another proof can be based on the generic
theorem of McNaughton and Zalestein, which states that every
finite semigroup of matrices over a field is torsion \cite{mac}.
We have given here a self contained proof for nonnegative integer
matrices.

The meaning of the condition (\ref{unbound}) in terms of cascade
graphs can be seen from the following simple example. If one
matrix in $\Sigma$ has those three entries (and no other) equal to
one, then we have two infinite and separate paths: one is a
circuit passing through the node $i$, the other is a circuit
passing through the node $j$. Those cycles are linked in a unique
direction, so that the first one is a source and the second one is
a sink, that eventually collects all these paths, as shown on
Figure \ref{fig-poly}.
\begin{figure}
 \centering
\includegraphics[width=0.3\textwidth ]{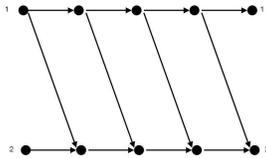}
\caption{A cascade graph with linear growth.} \label{fig-poly}
\end{figure}

We now prove that the criterion of Proposition
\ref{prop-unbounded} can be checked in polynomial time.\\

\begin{theo}\label{check-linear}
There is a polynomial time algorithm for verifying product
boundedness of families of nonnegative integer matrices.
\end{theo}

\begin{proof}
Assume we are given a finite set of nonnegative integer matrices
$\Sigma$. First, we decide between the cases $\rho = 0, \rho = 1$
and $\rho
> 1$ with the algorithm provided in the previous section.
In the first case $\max_t (\Sigma )$ is bounded, in the latter
it is not. The main problem is to check boundedness for the case
$\rho = 1$. By Proposition \ref{prop-unbounded} it suffices to
check if there exists a product $A \in \Sigma^*$ possessing the
property of equation \ref{unbound} for some indices $i, j$.
Consider the product graph $G^3$ with $n^3$ nodes defined as
follows. The nodes of $G^3$ are ordered triples $(i, j, k)$, where
$i, j, k \in \{1, \ldots , n\}$. There is an edge from a vertex
$(i, j, k)$ to a vertex $(i', j', k')$ if and only if there is a
matrix $A \in \Sigma$, for which $(A)_{i, i'}, (A)_{j, j'},
(A)_{k, k'} \ge 1$. (The adjacency matrix of $G^3$ is obtained by
taking the $3$-th Kronecker power of each matrix in $\Sigma$, and
by taking the maximum of these matrices componentwise.) The above
condition means that there are indices $i \ne j$ such that there
is a path in $G^3$ from the node $(i,i,j)$ to the node $(i,j,j)$.
The algorithm involves checking $n(n-1)$ pairs, and for each pair
at most $n^3$ operations to verify the existence of a path from
$(i,i,j)$ to  $(i,j,j)$. In total one needs to perform $n^5$
operations to check boundedness.
\end{proof}
\end{section}


\begin{section}{The rate of polynomial growth} \label{poly}
We have provided in the  previous section a polynomial time
algorithm for checking  product boundedness of sets of
nonnegative integer matrices. In this section we consider sets of
matrices that are not product bounded and we analyze the
 \emph{rate of growth} of the value $\max_t
(\Sigma )$ when $t$ grows. When the set $\Sigma$ consists of only
one matrix $A$ with spectral radius equal to one, the norm of
$A^k$ increases polynomially with $k$ and the degree of the
polynomial is given by the size of the largest Jordan block of
eigenvalue one. A generalization of this for several matrices is
given in the following theorem.\\

\begin{theo}\label{theo-poly}
For any finite set  $\Sigma $ of integer nonnegative matrices with
${\rho(\Sigma)=1}$ there are positive constants $C_1$ and $C_2$
and an integer $k \ge 0$ (the rate of growth) such that
\begin{equation}\label{c1c2}
C_1 t^k \ \le \ \max_t (\Sigma) \ \le \ C_2 t^k
\end{equation}
for all $t$. The rate of growth $k$ is the largest integer
possessing  the following property:
 there exist $k$ different ordered pairs of indices
 $(i_1,j_1), \dots , (i_k,j_k)$ such that for every pair $(i_s, j_s)$
there is a product $A\in \Sigma^{\star}$, for which
\begin{equation}\label{111}
 A_{i_s,i_s},  \ A_{i_s,j_s}, \ A_{j_s,j_s}\ \geq \ 1,
 \end{equation}
  and for each $1\leq s \leq k-1$,
there exists $B\in \Sigma^{\star}$ such that $B_{j_s,i_{s+1}}\geq
1$.
\end{theo}
The idea behind this theorem is the following: if we have a
polynomial growth of degree $k$, we must have a combination of $k$
linear grows that combine themselves successively to create a
growth of degree $k$.   This can be illustrated by the cascade
graph in Figure \ref{fig-superlin}.
\begin{figure}
 \centering
\includegraphics[width=0.3\textwidth ]{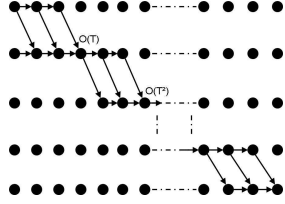}
 \caption{A cascade graph with polynomial growth.}
\label{fig-superlin}
 \end{figure}

Before we give a proof of Theorem \ref{theo-poly} let us observe
one of its corollary. Consider the ordered chain of maximal
strongly connected subsets $V_1, \ldots , V_I$ for our set
$\Sigma$. By Corollary \ref{corones} the elements $i_s, j_s$ of
each  pair $(i_s, j_s)$ belong to different sets, with, if $i_s
\in V_{i_s},j_s \in V_{j_s}$, $j_s > i_s$.  This implies that
there are less such couples than strongly connected subsets, and
then:\\

\begin{corr}\label{cork}
The rate of growth $k$ does not exceed $I-1$, where $I$ is the
number of strongly connected sets
of the family $\Sigma$. In particular, $k \le n-1$.
\end{corr}

We may now provide the proof of Theorem \ref{theo-poly}.

\begin{proof}
We shall say that a  node $i$ is $O(t^k) $ if there is a constant
$C>0$ such that $\, \max\limits_{A\in \Sigma^{t},1 \leq j\leq n}
A_{i,j} \le Ct^k$ for all $t$. Suppose that for some $k$ we have
$k$ pairs $(i_1,j_1), \dots , (i_k,j_k)$ satisfying the assumption
of the theorem.  We construct a cascade graph similar to the one
represented in Figure \ref{fig-superlin}: Let $A_s, \ s = 1,
\ldots , k$ and $B_s, \ s = 1, \ldots , k-1$ be the corresponding
products and $m$ be their maximal length. Then for any $s$ and any
$p \in \mathbb N$ one has $(A_s^p)_{i_s j_s} \ge p$, and therefore
$\bigl( A_1^{p}B_1 A_2^{p}B_2\dots A_k^{p}\bigr)_{i_1,j_{k}} \geq
p^k$ for any $p$. Denote this product by $D_p$ and its length by
$l_p$. Obviously $ l_p \le (pk + k-1)m$. For an arbitrary $t >
(2k-1)m$ take the largest  $p$ such that $l_p < t$. It follows
that $l_p \ge t- km$, and therefore $p \ge \frac{l_p}{km} - 1 +
\frac{1}{k} \ge \frac{t}{km} - 2 + \frac{1}{k}$. In order to
complete the product, take for instance $A_k^{t- l_p }$. Then the
product $D_p A_k^{t-l_p}$ has length $t$ and its $(i_1j_k)$-entry
is bigger than$ \ p^k \ge \bigl(\frac{t}{km} - 2 +
\frac{1}{k}\bigr)^k$, which is bigger than $Ct^k$ for some
positive constant $C$. This proves sufficiency. \\
It remains to establish the converse: if for some $k$ there is a node  that is
not $O(t^{k-1})$, then there exist $k$ required pairs of indices.
We prove this by induction in the dimension $n$ (number of nodes).
For $n = 2$ and $k=1$ it follows from Proposition
\ref{prop-unbounded}. For $n=2$ and $k>2$ this is impossible,
since one node (say, node $1$) is an invariant by Corollary
\ref{correduc}, then the edge $(1,2)$ is forbidden, and there is
at most $t+2$ paths of length $t$ (if all other edges occur at
each level).\\ Suppose the theorem holds for all $n' \le n-1$. Let
a node $i_0$ be not $O(t^{k-1})$.  Assume first that there are two
nodes $i, j$ of the graph $G(\Sigma)$ that are not connected by
any path. Therefore there are no paths containing both these
nodes. Hence one can remove one of these nodes (with all
corresponding edges) so that $i_0$ is still not $O(t^{k-1})$. Now
by induction the theorem follows. It remains to consider the case
when any pair of nodes is (weakly) connected. Take the
decomposition in strongly connected subsets $V_1, \ldots , V_I$
for $\Sigma$.  The
 nodes are ordered so that all the matrices in
$\Sigma$ are in  block upper triangular form. Let $p$ be the
smallest integer such that all nodes in $ G_p =\cup_{s=p}^IV_s$
are $O(1)$, \emph{i.e.}, $G_p$ is the biggest invariant on which
the number of paths is bounded. By Corollary \ref{corones} such
$p$ does exist. On the other hand, by the assumption we have $p
\ge 2$. Since the products in $\Sigma^{\star}$ restricted to the
subspace corresponding to $G_{p-1} = G_p \cup V_{p-1}$ are
unbounded, it follows from Proposition \ref{prop-unbounded} that
there is a pair $(i_k, j_k) \in G_{p-1}$ realizing equation
\ref{unbound}. Observe that $i_k \in V_{p-1}$ and $j_k \in G_p$.
Otherwise both these nodes are either in $V_{p-1}$ (hence the
restriction of $\Sigma^{\star}$ to $V_{p-1}$ is unbounded, which
violates Corollary \ref{corones}) or in $G_p$ (contradicts the
boundedness of $\Sigma^{\star}$ on $G_{p}$). Now consider the
products restricted on the set $ \cup_{s=1}^{p-1}V_s$. We claim
that at least one node is not $O(t^{k-2})$ in this restriction:
For any product in $\Sigma^{\star}$ of length $t$ consider the
corresponding cascade graph. Any path of length $t$ starting at a
node $i\in  \cup_{s=1}^{p-1}V_s$ consists of 3 parts (some of them
may be empty): a path $i \to v \in
 \cup_{s=1}^{p-1}V_s$ of some length $l$, an edge $v\to u \in G_{p}$,
and a path from $u$ inside $G_p$ of length $t-l-1$. Suppose that
each entry in the restriction of the products to $
\cup_{s=1}^{p-1}V_s$ is $O(t^{k-2})$, then for a given $l$ there
are at most $Cl^{k-2}$ paths for the first part ($C>0$ is a
constant), for each of them the number of different edges $v\to u
$ (counting with edges) is bounded by a constant $K$, and the
number of paths from $u$ to the end is bounded by $C_0$ by the
assumption. Taking the sum over all $l$ we obtain at most
$\sum_{l=0}^t CKC_0l^{k-2} = O(t^{k-1})$ paths, which contradicts
our assumption.\\  Hence there is a node in $ \cup_{s=1}^{p-1}V_s$
that is not $O(t^{k-2})$. Applying now the inductive assumption to
this set of nodes we obtain $k-1$ pairs $(i_s, j_s), \ s = 1,
\ldots , k-1$ with the required properties. Note that they are
different from $(i_k, j_k)$, because $j_k \in G_p$. It remains to
show that there is a path in $G(\Sigma)$ from $j_{k-1}$ to $i_k$.
Let us remember that $i_{k} \in V_{p-1}$. If $j_{k-1} \in V_{p-1}$
as well, then such a path exists, because $V_{p-1}$ is strongly
connected. Otherwise, if $j_{k-1} \in V_{j}$ for some $j < p-1$,
then there is no path from $i_k$ to $j_{k-1}$, which yields that
there is a path from $j_{k-1}$ to $i_k$, since each pair of nodes
is weakly connected.\end{proof}

 Let us note that the products of maximal growth constructed in the
proof of Theorem \ref{theo-poly}
 are not periodic, that is, the optimal asymptotic product is not the
power of one product. Indeed, we multiply the first matrix $A_1$
$p$ times, and then the second one $p$ times, etc.  This leads to
a family of products of length $t$ that are not the repetition of
a period.  In  general, those aperiodic products can be the
optimal ones, as illustrated by the following simple example.
\begin{equation}\nonumber
\Sigma \ = \ \left\{ \begin{pmatrix} 1&1&0 \\ 0&1&0\\
0&0&0\\\end{pmatrix}\, , \
\begin{pmatrix}
0&0&0\\ 0&1&1 \\ 0&0&1\\ \end{pmatrix}\right\}.
 \end{equation}
 Any finite product of these matrices has spectral radius equal to one
and has at most
 linear growth. Indeed, every $A \in \Sigma$ has rank at most two,
 therefore the condition of Theorem \ref{theo-poly} for any $k =2$ is
not satisfied for the product  $A$.
 Nevertheless, the aperiodic sequence of products of the
 type $A_1^{t/2}A_2^{t/2}$ gives a quadratic growth in $t$.
 It is interesting to compare this phenomenon with the well-known
finiteness property
 of linear operators \cite{cfbousch,cfblondel,cfkoz}: for this set of
matrices, the maximal behavior is a quadratic growth, which is
possible only for aperiodic products. On the other hand,
considering the boundedness of the products such phenomenon is
impossible: by Proposition \ref{prop-unbounded} if $\max_t
(\Sigma)$ is unbounded, this unbounded growth can always be
obtained by a periodic sequence. This fact is true only for
nonnegative integer matrices, since the following example gives a
set of complex matrices for which the products are unbounded while
all periodic products are bounded:
\begin{equation}\nonumber
\Sigma \ = \ \left\{ \begin{pmatrix} e^{i\theta 2\pi}&1 \\ 0&1\\
\end{pmatrix},\begin{pmatrix}
e^{i\theta 2 \pi}&0 \\ 0&1\\
\end{pmatrix}\right\}.
 \end{equation}
If $ 0 \leq \theta \leq 1$ is irrational, then the powers of any
$A \in \Sigma^*$ are bounded, while $\max_{t}(\Sigma )$ grows
linearly in $t$.\\

\begin{propp}
The rate of growth of a set of nonnegative integer matrices can be
found in polynomial time.
\end{propp}
\begin{proof}
For each pair $(i,j)$ of vertices one can check in polynomial time
whether there is a product $A$ such that $A_{i,j}\geq
1,\,A_{i,i}=A_{j,j}=1 $. For each couple of those pairs
$(i_1,j_1),(i_2,j_2)$, we can check in polynomial time whether
there is a path from $j_1$ to $i_2$, or from $j_2$ to $i_1$.
Finally we are left with a directed graph whose nodes are the
couples $(i,j)$ satisfying equation \ref{unbound} and with an edge
between the nodes $(i_1,j_1),(i_2,j_2)$ if there is a path from
$j_1$ to $i_2$. This graph is acyclic (because if there is also a
path from  $j_2$ to $i_1$ then there are two paths from $i_1$ to
itself, and $\rho >1$ by Lemma \ref{lemdiag}), and it is known
that the problem of finding a longest path in a directed acyclic
graph can be solved in linear time.
\end{proof}
\end{section}


\begin{section}{Polynomial growth for arbitrary
matrices}\label{problems} Theorem \ref{theo-poly} shows that for a
finite family $\Sigma$ of nonnegative integer matrices with joint
spectral radius equal to one the value $\max_t (\Sigma)$ is
asymptotically equivalent to $t^k$, where $k$ is an integer.
Moreover, we have shown that the exponent $k$ can be computed in
polynomial time. A natural question arises: do these properties
hold for all sets  matrices (without the constraint of nonnegative
integer entries)?\\

\begin{probb}\label{probl1}
Is this true that for any family of matrices $\Sigma$ (real or
complex) with $\rho (\Sigma ) = 1$ one has  $\max_t (\Sigma)
\asymp t^k$ for some integer $k$ ?
\end{probb}
In other words, is the asymptotic behavior of the value $\max_t
(\Sigma)$ really polynomial with an integer rate of growth? This
property can obviously be reformulated without  the restriction
$\rho (\Sigma ) = 1$ as follows: is it true that for any family of
matrices $\Sigma$ we have
\begin{equation}\label{prob1}
\max_t (\Sigma) \ \asymp \ \rho^t t^k,
\end{equation}
where $\, \rho \, = \, \rho (\Sigma )$ and $k$ is an integer ? A
more general problem arises if we remove the strict requirements
of asymptotic equivalence up to a positive constant:\\

\begin{probb}\label{probl2}
Is this true that for any family of matrices $\Sigma$ the
following limit
\begin{equation}\label{prob2}
\lim\limits_{t \to \infty} \frac{\ln \, \rho^{-t}\max_t
(\Sigma)}{\ln t},
\end{equation}
 exists and is always an
integer?
\end{probb}
In particular, does property (\ref{prob1}) or, more generally,
property (\ref{prob2}) hold for nonnegative integer matrices ? If
the answer is positive, can the rate of growth be computed?  We
have solved these problems only for the case $\rho = 1$.  Thus, is
it possible to obtain a sharper information on the asymptotic
behavior of the value $\max_t (\Sigma)$ as $t \to \infty$ than the
well-known relation $\lim\limits_{t \to \infty}\ln{ \max_t
(\Sigma)}/  t \ = \ \ln{\rho (\Sigma)}$? The question is reduced
to the study of the value $r(t) = \rho^{-t}\max_t (\Sigma)$.  For
some special families of matrices this question has appeared in
the literature many times. S. Dubuc in 1986 studied it for a
special pair of $2\times 2$ matrices in connection with the rate
of convergence of some approximation algorithm \cite{Du}. In 1991
I. Daubechies and J. Lagarias \cite{DL} estimated the value $r(t)$
for special pairs of $n\times n$ matrices to get a sharp
information on the continuity of wavelets and refinable functions,
and their technique was developed in many later works (see
\cite{Ref} for references).  In 1990 B. Reznik \cite{R}
formulated several open problems on the asymptotic of binary
partition functions (combinatorial number theory) that were
actually reduced to computing the value $r(t)$ for special binary
matrices \cite{P3}. This value also appeared in other works, in
the study of various problems \cite{DR,CH,Rio}. For general
families of matrices very little is known about the asymptotic
behavior of $r(t)$, although  some estimations are available.
First, if the matrices from $\Sigma$ do not have a nontrivial
common invariant subspace, then $r(t) \asymp 1$, \emph{i.e.}, $C_1
\le \rho^{-t}\max_t(\Sigma ) \le C_2$ for some positive constants
$C_1, C_2$ \cite{BW,P1,wirth02generalized}. So, in this case the
answer to problem \ref{probl1} is positive with $k = 0$. This
assumption was relaxed for nonnegative matrices in \cite{P3}. It
was shown that if a family of nonnegative matrices is irreducible
(has no common invariant subspaces among the coordinate planes),
then we still have $r(t) \asymp 1$. For all other cases, if the
matrices are arbitrary and may have common invariant subspaces,
 we have only rough estimations. For the lower bound
  we always have
$r(t) \ge C$ \cite{P1}.
 For the upper bound, as it was independently shown in
\cite{DL} and \cite{Bell},   we have $r(t) \le Ct^{n-1}$.  This
upper bound was sharpened
 in the following way \cite{CH}. Let $l$ be the maximal integer such
that
 there is a basis in ${\mathbb R}^n$, in which all the matrices
 from $\Sigma$ get a block upper-triangular form with $l$ blocks.
 Then $r(t) \le Ct^{l-1}$. The next improvement was obtained in \cite{P4}.
 Let  $\Sigma = \{A_1, \ldots , A_N\}$ and each matrix $A_d \in \Sigma$ are in upper triangular form, with diagonal
blocks $B_d^1, \ldots , B_d^l$. Let $s$ be the total
number of indices $j \in \{1, \ldots , l\}$ such that $\rho
(B_1^j,\ldots ,  B_N^j) = \rho (\Sigma )$. Then $r(t) \le
Ct^{s-1}$. Thus, for an arbitrary family of matrices we have $C_1
\le \rho^{-t}\max_t(\Sigma ) \le C_2t^{s-1}$. To the best of our
knowledge this is the sharpest information about the asymptotic
behavior of $r(t)$ available thus far.
\end{section}


\begin{section}{Conclusion and remarks}
The results of this paper completely characterize finite sets of
nonnegative integer matrices with bounded products and with
polynomially growing products. Without any changes the results can
be applied to general sets of nonnegative matrices, if the values
of the entries between zero and one are forbidden. Unlike the
proofs, which are quite technical, the results are easily
implementable in algorithms. One question we are not addressing in
this paper is that of the  exact computation of the joint spectral
radius when $\rho>1$; but this problem is known to be NP-hard even
for binary matrices.  We also provide an example of two matrices
whose joint spectral radius is one but for which the optimal
asymptotic behavior is not periodic. This example may possibly
help for the analysis of the finiteness property that was
conjectured in \cite{bpj1} to hold for binary matrices. Finally,
in the last section we leave several open problems on possible
generalizations of these results for more general sets of
matrices.
\end{section}

\subsection*{Acknowledgement}
This work was carried out while the second author was visiting the
Department of Engineering and Mathematics of the Universit\'e
catholique de Louvain (Belgium). That author wishes to express his
thanks to the University for its hospitality.


\begin{thebibliography}{05}
\bibitem{bpj1}Vincent D. Blondel, Rapha\"el Jungers, and Vladimir
Protasov. On the complexity of computing capacity of codes that
avoid forbidden difference patterns. Submitted to publication.
\end{thebibliography}
\end{document}